\newcommand{\bq}{\begin{equation}} \newcommand{\eq}{\end{equation}}
\newcommand{\ba}{\begin{eqnarray}}
\newcommand{\ea}{\end{eqnarray}}

\newcommand{\nl}{\\ \nonumber}

\def\ifmath#1{\relax\ifmmode #1\else $#1$\fi}%
\let\ifmathx0
\def\fixmath{\def\ifmath{\noexpand\ifmathx}}%

\def\ovMZ{\ifmath{{\overline M}_Z}}
\def\MZ{\ifmath{M_Z}}

\def\ovMZ{\ifmath{{\overline m}_Z}}
\def\ovGZ{\ifmath{{\overline \Gamma}_Z}}

\def\gam{\ifmath{\gamma}}%
\def\Zo{\ifmath{\mathrm {Z}}}

\def\Rgf{\ifmath{\mathrm {R}^f_\gam}}

\def\RZfi{\ifmath{\mathrm {R_Z}^{fi}}}

\def\Ffin{\ifmath{\mathrm {F}_n^{fi}}}

\documentstyle[12pt,bezier,epsf,epsfig]{article}
\begin{document}

\thispagestyle{empty}
 \begin{flushleft}
DESY 97--218
\\
hep-ph/9711305
\\
November 1997
 \end{flushleft}
 
 \noindent
 \vspace*{0.50cm}
\begin{center}
 \vspace*{2.cm} 
{\huge 
The $Z$ Boson Resonance 
\vspace*{3mm}
\\
and Radiative Corrections\footnote{Invited lecture given
at the XXI School of Theoretical Physics of the University of Silesia,
Katowice, on ``Recent Progress in Theory and
Phenomenology of Fundamental Interactions'', 19--24 September 1997,
Ustr{\'o}n, Poland}
 \vspace*{2.cm}               
}


{\large 
Tord Riemann
}
\vspace*{0.5cm}
 
\begin{normalsize}
{\it
Deutsches Elektronen-Synchrotron DESY, Institut f\"ur Hochenergiephysik
\\ 
IfH Zeuthen, Platanenallee 6, D-15738 Zeuthen, Germany
}
\end{normalsize}
\end{center}
 
 \vspace*{1.5cm} 
 \vfill 

\begin{abstract}
The $Z$ line shape is measured at LEP with an accuracy at the
per mille level.
Usually it is described in the Standard Model of electroweak
interactions with account of quantum corrections.
Alternatively, one may attempt different 
model-independent approaches in order to extract quantities like mass
and width of the $Z$ boson.
If a fit deviates from that in the standard approach, this may give hints for
New Physics contributions.
I describe two model-independent approaches and compare their
applications to LEP data with the Standard Model approach.
\end{abstract}

\vfill


  
\section{Introduction}
From 1989 till 1995 about 18 millions of $Z$ bosons have been produced at LEP1
and about 200\,000 at SLC. 
Due to this, and due to the lack of direct hints for the existence of a Higgs
boson, the $Z$ boson and its interactions became for several years the
central theme of tests of the Standard Model
\cite{Glashow:1961,Weinberg:1967,Salam:1968rm,Glashow:1970st,Bouchiat:1972iq}, 
recently accompanied by the discovery of the $t$ quark at
the Tevatron \cite{Abe:1995hr,Abachi:1995iq}.

The predictions of the Standard Model depend on the particle masses, fermion
mixings \cite{Fritzsch:1997ks},
and one coupling constant.
A central role plays also the weak mixing angle which relates 
(i) the $Z$ boson field and the photon to the symmetry fields, 
(ii) the two coupling constants of the electroweak theory,
(iii) the $Z$ and $W$ mass ratio, 
(iv) the vector and axial vector couplings of the $Z$ to fermions.

The corresponding electroweak Standard Model relations,
modulo radiative corrections, are : 
\ba
Z &=& \cos\theta_w \, W^0 - \sin\theta_w \, B
\\
\gamma &=& \sin\theta_w \, W^0 + \cos\theta_w \, B
\\
\label{gse}
g~{\sin \theta_w} &\equiv& e = \sqrt{4\pi\alpha_{em}}, 
\\
a_{lept} &=& - 0.5
\\
v_{lept} &=& - 0.5  \left(1-4\sin^2\theta_w\right) 
\\
\label{Gmu}
G_{\mu}/\sqrt{2} &=& g^2/(8M_W^2)
\\
M_Z &\equiv& M_W / \cos \theta_w
\label{zwr}
\ea
From (\ref{gse}), (\ref{Gmu}), and (\ref{zwr}) one derives:
\ba
M_Z >  M_W =
\sqrt{\frac{\pi\alpha_{em}}{G_{\mu}\sqrt{2}}}~\frac{1}{\sin\theta_w} 
> 37.281 \, \,{\rm GeV} 
\ea
as an absolute (though model-dependent) lower limit for the gauge boson masses.
From Born unitarity considerations it was expected that the Fermi theory
should loose validity above a mass scale of about ${\cal O}(100)$ GeV. 
In 1973, the first observation of neutrino-induced weak neutral current events
was reported by the Gargamelle collaboration operating at CERN.
One event was observed for the reaction 
${\bar \nu}_{\mu} e^- \to {\bar \nu}_{\mu} e^-$ 
from 360000 ${\bar \nu}_{\mu}$ and
375000 ${\nu}_{\mu}$ scatterings with 3\% background \cite{Hasert:1973cr}.
Since $\sigma \sim s=2\, E_{\nu} \, m_{target}$ at low energies,
the process $\nu N \to \nu N$ is about several thousand times ($\sim
m_p/m_e$) more frequent. 
In fact, about one third of $\nu N$ scattering is NC mediated and was also
observed by that time \cite{Hasert:1973ff}.   
From the ${\bar \nu}_{\mu}e$ cross-section, the authors derived with
90\% CL:
$0.1 < \sin^2 \theta_w < 0.6$.
This corresponds to:
\ba
M_Z \sim 75 \cdots 125~ {\mathrm GeV}
\ea
staying well in the limits mentioned above.

In 1983, at the $p{\bar p}$ collider SPS (CERN) 
both the $Z$ \cite{Arnison:1983mk,Bagnaia:1983zx} and the $W$ bosons
\cite{Arnison:1983rp,Banner:1983jy} were discovered and their masses
could be determined 
at that time with an accuracy of several GeV; in 1986:
\ba
M_Z = 92.6 \pm
1.7 ~ {\mathrm GeV}
\ea

At the end of 1989 LEP1 and SLC started operation and dominated the precision
experiments for tests of the electroweak Standard Model for a decade.
This may be exemplified by quoting the following improvements of precision from
August 1989 \cite{Altarelli:1989gg} till October 1997 \cite{Altarelli:1997sk}:
\ba
1989 ~~~~~~ &\to& ~~~~~~ 1997
\nonumber \\
\nl
M_Z = 91.120 \pm 0.160 ~ {\mathrm GeV} ~~&\to&~~ 91.186\,7 \pm 0.002\,0   
~ {\mathrm GeV}  
\\
\sin^2\theta_w^{eff} = 0.233\,00 \pm 0.002\,30 ~~&\to&~~ 0.231\,52 \pm
0.000\,23 
\\
{m}_{t}^{pred} = 130 \pm 50  ~ {\mathrm GeV} ~~&\to&~~ m_t^{meas} =
175.6 \pm 5.5 
~ {\mathrm GeV}  
\\
M_H^{pred} >  ~ {\mathrm few~GeV} ~~&\to&~~ \geq 77  ~ {\mathrm GeV}
\\
\alpha_s(M_Z) = 0.110 \pm 0.010 ~~&\to&~~ 0.119 \pm 0.003
\ea

The few remarks on the past may remind you
that $Z$ physics in times before the advent of LEP were exciting, and may
also indicate that the operation of LEP1 gave a striking
experimental support for the simplest realistic version of a renormalizable
theory of weak interactions.
Maybe it is worth mentioning that in the early Seventies the
Standard Model was considered by many of us to be an extremely
complicated, very artificially looking model. Soon later, when
hundreds of alternatives were constructed, minds changed and now many
of us wonder that such a simple construct like the Standard
Model survives one precision test after the other. 
But this needs not go on -- future is unpredictable. 
The only thing we know for sure: Future will be exciting.
\section{The $Z$ line shape}
The $Z$ boson may be studied as a resonance at LEP from a measurement
of the cross-section 
\ba
e^+ e^- \to (\gamma, Z) \to {\bar f}f ( + n \gamma)
\ea
as a function of the beam energy.
The determinations of mass $M_Z$ and width $\Gamma_Z$ 
are dominated by hadron production in a small region around the peak:
$|\sqrt{s} - M_Z | < 3$ GeV.
The $Z$ is not a pure Breit-Wigner resonance.
We want to study a $2 \to 2$ process with intermediate $Z$, but have
also virtual photon exchange.
In addition, there are huge $2 \to 3$ contributions due to initial state
radiation (ISR) of photons and due to final state radiation (FSR).
Further, many virtual corrections are contributing as quantum
  corrections: 
vertex insertions,
self energy insertions,
box diagrams, and all their iterations.
\section{Real photonic corrections
\label{qed}
}
The QED corrections may be taken into account by the following
convolution formula (\cite{Bardin:1989cw,Bardin:1991de,Bardin:1995aa}
and references therein):
\ba
\sigma(s)
&=&\int\frac{ds'}{s}\sigma_0(s')\, \rho\left(\frac{s'}{s}\right)
~+~\int\frac{ds'}{s}\sigma_0^{int}(s,s')
\, \rho^{int}
\left(\frac{s'}{s}\right)
\label{sigqed2}
\ea
\begin{itemize}
\item
$\rho(s'/s)$ -- the radiator 
describes initial and final state radiation,
including leading higher order effects and soft photon exponentiation;
\item
$\sigma_0(s')$ -- the basic scattering cross-section, which is the
object of investigation.
\end{itemize}
The $\rho^{int}(s'/s)$ takes into account the 
initial-final state interference effects which
are comparatively small (a few per mille) near the $Z$ resonance 
but are bigger off the resonance, and
$\sigma_0^{int}(s,s')$ is a function similar to $\sigma_0(s')$, but suppressed
if $\rho^{int}(s'/s)$ is small.

The dominant part of the QED corrections is ISR:
\ba
\rho(s'/s) &=& \beta (1-s'/s)^{\beta-1}\delta^{soft+virtual}+\delta^{hard}
\ea
where 
\ba
\beta &=& 2\frac{\alpha_{em}}{\pi} (L-1), ~~~~ L = \ln(s/m_e^2) 
\\
\delta^{soft+virtual} &=&
1
+\frac{\alpha_{em}}{\pi} \left[\frac{3}{2} L +2\zeta(2) -2 \right]
+{\cal O} \left( \frac{\alpha_{em}}{\pi} \right)^2
\\
\delta^{hard} &=& -\frac{\alpha_{em}}{\pi} (1+s'/s) \left(L-1\right)
+{\cal O}
\left( \frac{\alpha_{em}}{\pi} \right)^2
\ea

It is impossible to perform a reasonable model-independent fit
to the $Z$ resonance shape without a dedicated treatment of QED
corrections.
Their influence on the resonance shape near the peak (at
LEP1 energies) will be discussed in the next section. 
Here I show that the so-called radiative tail is proportional to
$M_Z/\Gamma_Z$; at higher energies (e.g. at LEP2) this is substantial 
but may be influenced by experimental cuts.

If we apply no cuts on the photon kinematics (i.e. observe none of
them), the ${\cal O}(\alpha_{em})$ corrections to the pure $Z$ contribution
to $\sigma_{tot}$  are \cite{Bardin:1988ze}:
\ba
\nonumber
\sigma_{tot}^Z 
&=&
\frac{4}{3} \frac{\pi \alpha_{em}^2}{s} \left| \chi(s) \right|^2
\Biggl\{(v_e^2 + a_e^2)(v_f^2 + a_f^2)
\Bigl[ 1 + \frac{\alpha_{em}}{\pi} ( Q_e^2 H_0^T(s,m_Z) 
\\ && 
+~ Q_f^2 H_2^T(s,m_Z) )\Bigr] 
+ 4v_ea_ev_fa_f \left[\frac{\alpha_{em}}{\pi}Q_eQ_fH_4^T(s,m_Z) \right] 
\Biggr\}
\ea
with $R_Z=m_Z/s, m_Z = M_Z^2-iM_Z \Gamma_Z$ and $\chi$ defined in
(\ref{basic_defs}).   
The numerically most important contribution is ISR:
\ba
\nonumber
H_0 ^T(s,m_Z) 
&=&
\left( \frac{\pi^2}{3} - \frac{1}{2}\right) + 
\left( \ln \frac{s}{m_e^2}-1\right)
\Re e ~\Biggl[ 
2R_Z +\frac{1}{2} -|R_Z|^2
\\ && + \frac{i \cdot s}{M_Z\Gamma_Z}(1-R_Z^*) R_Z
(1+R_Z^2) \ln\frac{R_Z-1}{R_Z}
\Biggr]
\ea
The last term describes the radiative tail of the $Z$ resonance.
If $\Re e R_Z < 1$, i.e. if $M_Z < s$, it is:
\ba
\Re e~i \cdot \frac{M_Z}{\Gamma_Z}\ln\frac{R_Z-1}{R_Z} &=&
{\cal O}(\pi)\cdot  \frac{M_Z}{\Gamma_Z}
\ea
Otherwise (i.e. $R_Z > 1$, at energies below the peak) this term stays small.

The analytic structure of this QED contribution, as well as that of the
others, is completely different from any Born-like expression.
This will be of importance if one tries to describe measured
cross-sections by simple parameterizations: The QED corrections have
to be treated explicitly.
\section{Model (I):  A model-independent ansatz
\label{I}
}
QED corrections are treated by the convolution formula introduced in
section \ref{qed}.
For a careful discussion of their influence on height and location of
the $Z$ peak see \cite{Beenakker:1990ec}.
 
The following ansatz for $\sigma_0(s)$ is a good choice
without explicit reference to the Standard Model
\cite{Borrelli:1990bd,Jegerlehner:1991ed,Stuart:1991xk,Leike:1991pq}:
\ba
\sigma_0(s) =
\frac{4}{3} \pi \alpha_{em}^2
\left[ \frac{r^{\gamma}}{s} +
\frac {s\cdot R + (s - M_Z^2)\cdot J} 
{\left|s-M_Z^2 + i s \Gamma_Z/M_Z\right|^2}
\right]
\label{sigqed3}
\ea
The line shape is described by five parameters:
\begin{itemize}
\item $r^{\gamma} \sim \alpha_{em}^2(M_Z^2)$ -- may be assumed to be known
\item  $M_Z,~~ \Gamma_Z$
\item $R$ -- measure of the $Z$ peak height
\item $J$ -- measure of the $\gamma Z$ interference
\end{itemize}
A simpler, also reasonable ansatz would be a pure Breit-Wigner function:
\ba 
\sigma_0^{(Z)}(s) \sim
\frac {M_Z^2\cdot  R } {\left|s-M_Z^2 +i M_Z \Gamma_Z\right|^2}
\ea
The effects of the QED corrections are huge; among others, a shift of
the peak position arises:  
\ba
\sqrt{s_{\max}} - M_Z &=& 
\delta_{QED} =
\frac{\pi}{8} \beta 
\left(1+\delta^{soft+virtual}\right) \Gamma_Z
+ \,\mbox{small corr's.}
\nl
&\approx&~ 90 \,\mbox{MeV}
\ea
From the replacements $M_Z^2  R \to M_Z^2 s$, $M_Z \Gamma_Z \to (s/M_Z)
\Gamma_Z$ additional shifts arise:
\ba
\sqrt{s_{\max}} - M_Z &=& 
\delta_{QED} \oplus 
\frac{1}{4} \frac{\Gamma_Z^2}{M_Z} 
\ominus
\frac{1}{2}\frac{\Gamma_Z^2}{M_Z} \sim \left( 90+ 17 - 34\right)~~\mbox{MeV}
\ea
Finally, adding the effect of the $\gamma Z$ interference $J$: 
\ba
\sqrt{s_{\max}} - M_Z 
&=& 
\delta_{QED} \oplus 
\frac{1}{4} \frac{\Gamma_Z^2}{M_Z} \left(1+\frac{J}{R}\right)
\ominus
\frac{1}{2}\frac{\Gamma_Z^2}{M_Z}
\nl
&\sim& \left[90 + 17 \times \left(1+\frac{J}{R}\right) -34\right]
~~\mbox{MeV} 
\ea
Neglecting this interference (setting $J$=0) leads to an erroneous
systematic shift of the $Z$ mass of 17~MeV$\otimes(J/R)$.
If one wants to take into account the $J$, a model for its
prediction is needed.
In the Standard Model (\cite{Clare:1997AA}, table 7):
$J=0.22, R=2.96$ for hadron production. Thus, 
$J/R~ \otimes$ 17 MeV $\approx$ 1.2 MeV.
\subsection{$Z$ line shape fit (I)}
With the model-independent ansatz, the following nearly uncorrelated
observables may be determined from the $Z$ peak data
\cite{DRWard:1997AA,Clare:1997BB}: 
\ba
M_Z &=&  91.186\,7 \pm 0.002\,0~~ \mbox{GeV}~~~~(\delta=0.0025~\%)
\\
\Gamma_Z &=& 2.494\,8 \pm 0.002\,5~~ \mbox{GeV}~~~~(\delta=1.3~\%)
\\
\sigma_0^{had} &=& 41.486 \pm 0.053~~{\mbox{nb}}~~~~(\delta=1.9~\%)
\\
R_l = \frac{\sigma_0^{had}}{\sigma_0^{lept}} &=& 20.775 \pm
0.027~~~~(\delta=1.5~\%) 
\\
A_{FB,0}^{lept} &=& 0.017\,1 \pm 0.001\,0
\ea
Here, $M_Z, \Gamma_Z, \sigma_0^{had}$ are from $\sigma^{had}(s)$,
while $R_l$ and $A_{FB}$ from $\sigma^{lep}(s)$:
with $\sigma_0^{had(lep)}$ as hadronic (leptonic) peak cross-section, and
$A_{FB,0}^{lept}$ as forward-backward asymmetry at the peak.
These parameters are considered to be primary parameters in contrast
to derived ones, e.g. the effective
leptonic weak neutral current couplings of leptons or the effective
weak mixing angle  \cite{DRWard:1997AA,Clare:1997BB}: 
\ba
v_l &=& - 0.037\,93 \pm 0.000\,58 
\\
a_l &=& - 0.501\,03 \pm 0.000\,31
\\
\sin^{2}\theta_{w}^{eff} 
\equiv \frac{1}{4}\left( 1-\frac{v_l}{a_l}\right)
&=& 0.231\,52 \pm 0.000\,23
\label{sw1}
\ea
\section{Model (II): Virtual corrections in the Standard Model}
All virtual corrections may be written  in some theory,
e.g. the Standard Model, for massless particle production in the
following way (see
e.g. \cite{Bardin:1989di,Bardin:1992jc2,Bardin:1995a3} and references
therein): 
\ba 
\nonumber
{\bf\cal M}_{net} &\sim&
\frac{\alpha_{em}}{s} \Biggl\{
\frac{\alpha_{em}(s)}{\alpha_{em}}\left|Q_e Q_f \right| \gamma_{\beta} \otimes
\gamma_{\beta}  
+ \chi(s) \varrho_{ef} 
\Bigl[ L_{\beta} \otimes L_{\beta} 
\nl && -~ 4 s_w^2 |Q_e| \kappa_e \gamma_{\beta} \otimes L_{\beta} 
- 4 s_w^2|Q_f|  \kappa_b L_{\beta} \otimes \gamma_{\beta}
\\ && +~
16 s_w^4 |Q_eQ_f| \kappa_{eb} \gamma_{\beta} \otimes \gamma_{\beta}
\Bigr] 
\Bigr\}
\label{net}
\ea
We use short notations: $L_{\beta} = \gamma_{\beta}(1+\gamma_5 )$, 
$A_{\beta} \otimes B_{\beta} = [ \bar v_e A_{\beta} u_e ]
                        \cdot [ \bar u_b B_{\beta} v_b ]$, 
and
\ba
\chi = 
\chi(s) =  \frac{G_{\mu}}{\sqrt{2}} \frac{M_Z^2}{8\pi\alpha_{em}}
\frac{s}{s-M_Z^2+iM_Z\Gamma_Z(s)} 
, ~~
\Gamma_Z(s) = \frac{s}{M_Z^2} \Gamma_Z 
\label{basic_defs}
\ea
The effective Born cross-section now is uniquely determined once the
net matrix element ${\bf\cal M}_{net}$ is known:
\ba
\sigma_0(s)
&=&
N_c^f \sqrt{1-4m_f^2/s} ~~\frac{4\pi\alpha_{em}^2}{3s} \times
\nl
&& \Biggl\{
\Bigl(1+\frac{2 m_f^2}{s}\Bigr) 
\Bigl[
|Q_eQ_f|^2 \frac{|\alpha_{em}(s)|^2}{\alpha_{em}^2} 
+ 2 |Q_eQ_f| \Re e \Bigl( \chi  
\frac{\alpha_{em}^*(s)}{\alpha_{em}} \varrho_{ef} v_{ef} \Bigr)
\nl && +~
|\chi\varrho_{ef}|^2  (1 + |v_e|^2 + |v_f|^2+ |v_{ef}|^2 
)
\Bigr] 
-  \frac{6m_f^2}{s} |\chi\varrho_{ef}|^2 (1 + |v_e|^2)
 \Biggr\}
\label{sigeff0}
\ea
with
\ba
v_{i} &=& 1 - 4 s^2_w |Q_i|\kappa_i, ~~~~i=e,f
\\
v_{ef} &=& 1 - 4 s^2_w |Q_e| \kappa_e 
- 4 s^2_w |Q_f| \kappa_f + 16 s^4_w |Q_eQ_f| \kappa_{ef}
\ea
Further, $N_c^f=1, 3$ is the colour factor and QCD corrections also
have to be taken into account. 

The virtual corrections with higher order parts are (see for details
\cite{Bardin:1989di,Bardin:1995a3,Bardin:1998AB} and references therein): 
\ba
\varrho_{ef} &=& 
\frac{\left(1 + \tau_f\right)^2} 
{ 1 - \Delta \rho  - 
\Delta \varrho_{ef}^{\mathrm{1loop},\alpha}
+ 
\Delta {\bar \rho}^{\alpha}
+ 
2 \Delta{\bar \rho}_f
- 
\Delta \rho^{\mathrm{2loop},\alpha \alpha_s} + 
\Delta {\bar
  \rho}^{\alpha \alpha_s} 
} 
\label{rofs}
\\
\kappa_f &=& 
\Biggl[ 1 + 
\left(
  \Delta \kappa_f^{\mathrm{1loop},\alpha}
- \frac {c_w^2} {s_w^2} (\Delta {\bar \rho}^{\alpha} + {\bar X})
+ \Delta{\bar \rho}_f
\right)
\nl
&&+~ 
\left(
   \Delta \kappa^{\mathrm{2loop},\alpha \alpha_s}
- \frac {c_w^2} {s_w^2} \Delta {\bar \rho}^{\alpha \alpha_s}
\right) \Biggr]
\frac{ 1 + \frac {\displaystyle c_w^2}{\displaystyle s_w^2} ( \Delta
  \rho + X ) } 
{1+\tau_f} 
\label{kapafs}
\\
\kappa_{ef} &=& 
\Biggl[1 + 
\left(
\Delta \kappa_{ef}^{\mathrm{1loop},\alpha}
- 2 \frac {c_w^2} {s_w^2} (\Delta {\bar \rho}^{\alpha} + {\bar X} ) 
+ 2 \Delta{\bar \rho}_f
\right)
\nl &&+~ 2 \left(
\Delta \kappa_f^{\mathrm{2loop},\alpha \alpha_s}
- \frac {c_w^2} {s_w^2} \Delta {\bar \rho}^{\alpha \alpha_s}
\right)
\Biggr]
\frac{
\left(1 + \frac {\displaystyle c_w^2}{\displaystyle s_w^2} 
( \Delta \rho + X ) 
\right)^2}
{\left( 1+\tau_f\right)^2} 
\ea
The corrections $\Delta{\bar \rho}_f $ and 
$\tau_f $ contribute only to $b$ quark pair production.

The one loop form factors are, due to the $ZZ$ and $WW$ box contributions,
dependent on the scattering angle $\vartheta$: 
\ba
\rho^{\rm{1loop},\alpha}_{ef}(s,\cos\vartheta)
&=&
1 + \Delta \rho_{ef}^{non-box}(s)
+ \rho^{\rm{box}}_{ef}(s,\cos\vartheta)
\nl &&
+~ \delta_{fb} 
\biggl[\delta\rho^{t}_{eb}(s) 
+\delta\rho^{{\rm{box}},t}_{WW}(s,\cos\vartheta)\biggr]
\label{rhos}
\\
\kappa^{\rm{1loop},\alpha}_e(s,\cos\vartheta)
&=&
\kappa(e,f) + \delta_{fb}
\left[
\delta\kappa^t_{e}(s)
-\delta\rho^{{\rm{box}},t}_{_{WW}}(s,\cos\vartheta)
\right]
\\
\kappa^{\rm{1loop},\alpha}_f(s,\cos\vartheta)
&=&
\kappa(f,e) - \delta_{fb} \left[ \delta\rho^t_{eb}(s)
+\delta\rho^{{\rm{box}},t}_{_{WW}}(s,\cos\vartheta)
\right]
\ea
with
\ba
\kappa(e,f) &=&
1 + 
\Delta \kappa^{non-box}(e,f;s)
+ \kappa^{\rm{box}}_{e}(s,\cos\vartheta)
\label{kappaes}
\ea
and finally
\ba
\kappa^{\rm{1loop},\alpha}_{ef}(s,\cos\vartheta)
&=&
1 + 
\Delta \kappa_{ef}^{non-box}(s)
+ \kappa^{\rm{box}}_{ef}(s,\cos\vartheta)
\nl &&
-~ \delta_{fb}\biggl[\delta\rho^{t}_{eb}(s) 
+\delta\rho^{{\rm{box}},t}_{_{WW}}(s,\cos\vartheta)\biggr]
\label{kappaefs}
\ea
The form factors with super index $t$ are for $b$ quark pair
production only.
If the total cross-section is written in terms of four form factors with a
dependence on only $s$, it is implicitly assumed that the dependence of the
weak box terms on $\cos\vartheta$ is negligible.
This is a very good approximation near the $Z$ resonance but not at
much higher energies. 

Further, we need expressions for $M_W$, $\Gamma_Z$, $\alpha_{em}$ and
a reasonable treatment of QCD corrections and explicit expressions for the
structures shown above.
The $W$ boson mass is:
\ba
M_{W}=M_{Z}\sqrt{1
-\sqrt{1-\frac{4\pi\alpha_{em}}{\sqrt{2}G_{\mu}M_Z^2[1-\Delta r]}}} 
\label{wmass}
\ea
with
\ba
\frac{1}{1-\Delta r} &=&
\frac{1} {\left(1 - \Delta \alpha_{em} \right)
\left( 1 +{\displaystyle \frac {c^2_{w}}{s^2_{w}}} (\Delta \rho + X) \right)
- \Delta r_{\mathrm{rem}} }
\label{romus}
\ea
and
\ba
\Delta r_{\mathrm{rem}} &=&
\Delta r^{\mathrm{1loop},\alpha}
+ \Delta r^{\mathrm{2loop},\alpha \alpha_s}
+ \frac{c^2_{w}}{s^2_{w}} 
\left(
\Delta {\bar \rho}^{\alpha} + \Delta {\bar \rho}^{\alpha\alpha_s} +
{\bar X} \right)
- \Delta \alpha_{em}
\label{drrem}
\ea
  
For the $Z$ width 
\cite{Akhundov:1986fc,Jegerlehner:1986vs,Beenakker:1988pv,Bernabeu:1988me}, 
$\alpha_{em}$ 
\cite{Eidelman:1995ny,Jegerlehner:1996ab}, as well as QCD corrections
\cite{Chetyrkin:1994js3}, and all the other expressions
left out here I have to refer to
literature quoted above and to references therein.  
\subsection{$Z$ line shape fit (II)}
I quoted all the above formulae in order to demonstrate explicitly how involved
a Standard Model fit ansatz is.
The input quantities are: $\alpha_{em}, G_{\mu}$ (for $M_W$), $M_Z, m_f, M_H,
\alpha_s$. 
Some of them are precisely known (e.g. $G_{\mu}$), others are subject of
determination at LEP (e.g. $M_Z$), others are completely unknown ($M_H$).  
The $t$ quark mass may be determined from weak loop corrections at LEP or
directly from $t$ quark production at Fermilab.

Quantities like the $Z$ width or the weak mixing angle are not a subject of
fits since they are considered to be secondary quantities.
In this respect, there is a basic difference to the approach of the foregoing
section. 

The most recent Standard Model fit is \cite{DRWard:1997AA,Clare:1997BB}.
The $t$ quark mass from Fermilab is:
\ba
m_t &=& 175.6 \pm 5.5~~ \mathrm{GeV}
\ea
The global fit to all data yields:
\ba
M_Z  &=& 91.186\,7 \pm 0.002\,0 ~~  \mathrm{GeV}
\\
m_t &=& 173.1 \pm 5.4  ~~\mathrm{GeV}
\\
M_H &=& 115 ^{+116}_{-66} ~~\mathrm{GeV}
\\
\alpha_s(M_Z) &=& 0.120 \pm 0.003
\ea
In a next step, one may calculate the other quantities like the $Z$ width and
relate to values from model-independent fits. 
Whatever one does, there is no unique hint to New Physics. 
For a detailed discussion of this see \cite{Altarelli:1997sk}. 
\section{Model (III): The S-matrix approach
}
In view of the extremely high precision in the measurement of mass and width of
the $Z$ boson one may question the ansatz used.
A potential bias in the Standard Model (if it is not correct) would not show up
in the experimental errors. 
For this reason, one is interested in a model-independent $Z$ peak description
with a minimum of assumptions.
We saw in sections \ref{qed} and \ref{I} that we must take into account real
photonic corrections; but we saw also that this is possible with the use of an
effective Born cross-section to be folded with a function depending on the
kinematics, $M_Z$, and $\Gamma_Z$.

When the $Z$ boson is treated as a resonance, the S-matrix
approach\footnote{For an introduction to S-matrix theory, see 
\cite{Eden:1966AA,Bohm:1994AA}.} may be used for its description.

This was proposed in the context of LEP physics in \cite{Stuart:1991xk}, where
the perturbation expansion in the Standard Model was studied.
In \cite{Leike:1991pq} it was proposed to use this approach for a direct fit to
LEP data and the first S-matrix fit was performed therein.
The first fit by a LEP collaboration was due to L3
\cite{Adriani:1993gk,Adriani:1993ca}. 
The treatment of asymmetries near the peak was discussed in
\cite{Leike:1991pq}. 
For the role of QED corrections to asymmetries see
\cite{Riemann:1992gv}. 

A recent survey on the definition of $Z$ mass and width and their treatment in
fermion pair production is \cite{Riemann:1997tj}.
Here, I give a short introduction to the technical essentials.
 
Consider 
four independent helicity amplitudes in the case of massless fermions $f$:
\ba
\label{eqn:mat0}
{\cal M}^{fi}(s) = \frac{\Rgf}{s} + \frac{\RZfi}{s-s_Z} +
\sum_{n=0}^\infty \frac{\Ffin}{\ovMZ^2} \left(\frac{s-s_Z}{\ovMZ}\right)^n 
,~~\  i=1,\ldots,4.
\ea
The position of the \Zo\ pole in the complex $s$ plane is given by 
$s_{\Zo}$:
\ba
\label{szb}
s_Z = \ovMZ^2 - i \ovMZ \ovGZ.
\ea
The \Rgf\ and \RZfi\ are complex residua of the photon and the \Zo\ boson,
respectively. One may approximate (\ref{eqn:mat0}) by setting $\Ffin \to 0$.
There are four residua \RZfi\ for
$e^-_Le^+_R \to f^-_L f^+_R, 
e^-_Le^+_R \to f^-_R f^+_L,
e^-_Re^+_L \to f^-_R f^+_L, 
e^-_Re^+_L \to f^-_L f^+_R$.
The amplitudes
${\cal M}^{fi}(s)$ give rise to four cross-sections $\sigma_i$:
\ba
\renewcommand{\arraystretch}{1.2}
\label{eqn:xs14}
\begin{array}{lclcl}
&&\sigma_{T}^0(s)      & = & +~  \sigma_0 + \sigma_1 + \sigma_2 + \sigma_3,
  \\
\sigma_{\mbox{\scriptsize \it lr-pol}}^0(s) & = &
\sigma_{FB}^0(s)     & = & +~  \sigma_0 - \sigma_1 + \sigma_2 - \sigma_3,
 \\
\sigma_{\mbox{\scriptsize \it FB-lr}}^0(s)& = &
\sigma_{pol}^0(s)    & = & -~  \sigma_0 + \sigma_1 + \sigma_2 - \sigma_3
, \\
\sigma_{lr}^0(s)& = &
\sigma_{\mbox{\scriptsize \it FB-pol}}^0(s) & = & -~  \sigma_0 -
\sigma_1 + \sigma_2 + \sigma_3.
\end{array}
\renewcommand{\arraystretch}{1.}
\ea
Here, the $\sigma_{T}^0$ -- the total cross-section, 
$\sigma_{FB}^0$ -- numerator of the forward-backward asymmetry, 
$\sigma_{pol}^0$ -- that of the final state polarization etc.
All these cross-sections may be parameterized by the following master
formula ($A=T,FB, \ldots$): 
\ba
\renewcommand{\arraystretch}{2.2}
\label{eqn:smxs}
\sigma_A^0(s)
&=&
\displaystyle{\frac{4}{3} \pi \alpha_{em}^2
\left[ \frac{r^{\gamma f}_A}{s} +
\frac {s r^f_A + (s - \ovMZ^2) j^f_A} {(s-\ovMZ^2)^2 + \ovMZ^2 \ovGZ^2}
\right]} + \ldots
\renewcommand{\arraystretch}{1.}
\ea
The parameters are related to the residua of the pole terms. 
The $r^{\gamma f}_A$ is the photon exchange term and assumed to be known.
It vanishes for all asymmetric cross-sections.
The $Z$ exchange residuum $r^f_A$ and the \gam\Zo-interference
$j^f_A$ are, together with $Z$ mass and width, subject of a fit.

Without QED corrections, asymmetries are:
\ba
\label{eqn:mi_asy}
{\cal A}_A^0(s) = \frac{\sigma_A^0(s)}{\sigma_{T}^0(s)}
=
A_0^A + A_1^A \left(\frac{s}{\ovMZ^2} - 1 \right) + \ldots,~~~ A \neq T
\ea
They take the above extremely simple approximate form around the \Zo\
resonance. 
At LEP1, the higher order terms in the Taylor expansion may be neglected since
$(s/\ovMZ^2-1)^2  < 2 \times 10^{-4}$.
The coefficients have a quite simple form:
\ba
\label{eqn:a0}
A_0^A 
= \frac{r_A^f} {r_T^f}, ~~~~
A_1^A =
\left[ \frac{j^f_A}{r_A^f} -  \frac{j_T^f}{r_T^f} \right] A_0^A
\ea

The variation of asymmetries with $s$ near the peak is due to the $\gamma Z$
interference ($A_1$). 
QED corrections modify the coefficient $A_1$, but in a model-independent form.
It is important to note that they may not be neglected in a fit due to 
resonance tail effects similarly as was discussed above for the cross-section.

Another comment is necessary concerning the definition of mass and width of the
$Z$ boson.
The so-called pole definition with a constant width (\ref{szb})
as a natural consequence of
the S-matrix ansatz leads to different numerical values compared to
the usual Standard Model approach (\ref{basic_defs})
\cite{Bardin:1988xt,Berends:1988bg,Argyres:1995ym}.
A very precise approximation is:
\ba
\label{mmggzz}
\ovMZ & = & [ 1 + (\Gamma_Z/M_Z)^2 ]^{-\frac{1}{2}} \MZ
\approx 
M_Z - \frac{1}{2} \Gamma_Z^2/M_Z = M_Z -34~{\mbox{MeV}}
\label{eqn:mbar}
\ea
Similar observations were made in 1968 for hadron resonances
\cite{Gounaris:1968mw}. 
\subsection{$Z$ line shape fit (III)}
The interest of the community in an S-matrix based fit to the LEP data has
several origins.
One is the wish for a model-independent description of the resonance. 
Closely related is the question on the number of independent parameters needed
to describe the peak:
four (per channel) suffice to describe a cross-section: $M_Z, \Gamma_Z, r_T,
j_T$, provided we assume QED interactions to be understood. Among these
parameters, $Z$ mass and width are universal for all channels.
Any asymmetry introduces two additional degrees of freedom (per
channel): $r_A, j_A$. 

There are practical aspects of all this.
If the number of different energy points needed for a scan of the $Z$ peak is
asked for, the answer is at least five (four plus one) for cross-sections, 
at least three (two plus one) for asymmetries.
Further, the $\gamma Z$ interferences $j_A$ form separate  degrees of freedom. 
The $j_T$ and $M_Z$ are highly correlated. 
This became more important recently when the highest
statistics were taken, and also with the data collected at energies farer
away from the peak. 
There the interference becomes more influential.           

Recent experimental studies are summarized in \cite{Clare:1997AA}.
The data of table~\ref{table1} are obtained from the $Z$ line
shape scans at LEP which were performed mainly in 1993 and 
1995 (from table~7 of~\cite{Clare:1997AA}\footnote{Note that the table
  shows values of the complex pole mass $\ovMZ$.
The Standard Model fits use the on shell mass $M_Z$; the relation of both is  
given in (\ref{eqn:mbar}).   
}).
The biggest error correlations are shown in table~2 (from table~8
of~\cite{Clare:1997AA}). 
Including into the analysis cross-sections measured at TRISTAN energies
does not improve substantially e.g. the resolution of $M_Z$ and $j_T$
\cite{Clare:1997AA}.

\begin{table}[htbp]\centering
\begin{tabular}{|c|r@{$\pm$}l|r@{.}l|}
\hline
Parameter & \multicolumn{2}{c|}{S-matrix fit} & 
            \multicolumn{2}{c|}{SM Prediction}
\\
\hline
\hline
\ovMZ 
[GeV]      &  91.153\,4  & 0.003\,3  & \multicolumn{2}{c|}{--} 
\\
$\Gamma_Z$ [GeV] &   2.492\,4  & 0.002\,6  &  2&493\,2
\\
\hline
$r_T^{had}$      &   2.962\,3  & 0.006\,7  & 2&960\,3 
\\
$j_T^{had}$     &    0.15      & 0.15      &  0&22
\\
\hline
$r_T^{lept}$     &   0.142\,39 & 0.000\,34 &  0&142\,53
\\
$j_T^{lept}$     &   0.009     & 0.012  &  0&004
\\
\hline
$r_{FB}^{lept}$  &   0.003\,04 & 0.000\,18 &  0&002\,66
\\
$j_{FB}^{lept}$  &   0.789   & 0.013       &  0&799
\\  \hline
\end{tabular}

\vspace*{5mm}
\caption
{\it
Results from a combined LEP1 line shape fit
\label{table1}
}
\end{table}

\begin{table}[thbp]\centering
\begin{tabular}{|r@{--}l|r@{.}l|}
\hline
\multicolumn{2}{|c|}{Correlation} & \multicolumn{2}{c|}{Value}
\\
\hline
\hline
$M_Z $ & $ j_T^{had}$  & --0&77
\\
$M_Z $ & $ j_T^{lept}$ & --0&47
\\
$\Gamma_Z $ & $ r_T^{had}$ & 0&80
\\
$\Gamma_Z $ & $ r_T^{lept}$ & 0&62
\\
$r_T^{had} $ & $ r_T^{lept}$ & 0&78
\\
$j_T^{had} $ & $ j_T^{lept}$ & 0&49
\\  \hline
\end{tabular}

\vspace*{5mm}
\caption
{\it
Biggest correlations in the S-matrix fit
\label{table2}
}
\end{table}

\section{Summary}
We have presented three different approaches to a numerical
analysis of the $Z$ boson line shape -- a Breit-Wigner ansatz, the Standard
Model of electroweak interactions, and the S-matrix approach;
QED corrections must be applied. 
They all agree basically in the determination of the $Z$ mass.
The S-matrix approach treats the $\gamma Z$ interference as an
independent quantity, which enlarges the error for $M_Z$.
Two different mass definitions are used but may be related with a high
accuracy. 
In the Standard Model, $\Gamma_Z$ is a derived quantity;
the other two approaches allow direct fits.
The approaches agree numerically for the $Z$ width.

We see an essential difference of the S-matrix ansatz to the Standard Model.
The latter assumes {\em fixed} relations among many of the parameters. 
They rely thus on stronger theoretical assumptions. 

{\em From the strong experimental correlations in the S-matrix fit
  together with   the excellent 
agreement of the central values of fitted parameters in all fit scenarios one
may conclude that the different scenarios are highly compatible with each
other.}

\section*{Acknowledgements}
I would like to thank the organizers of the School, Karol Kolodziej and Marek
Zralek, for the invitation to present a lecture and for the generous
hospitality. 

\def\href#1#2{#2}
\bibliography{%
/home/phoenix/riemann/Bibliography/ca,%
/home/phoenix/riemann/Bibliography/radcorr,%
/home/phoenix/riemann/Bibliography/basics,%
}
\bibliographystyle{/home/phoenix/riemann/Bibliography/utphys_t}
\end{document}